# Highly anisotropic magnetic domain wall behavior in-plane magnetic films

Xiaochao ZHOU[1,2], Nicolas VERNIER[2,3], Guillaume AGNUS[2], Sylvain EIMER[2], Ya ZHAI[1]

1. School of Physics, Southeast University, 211189 Nanjing, China
2. Centre de Nanosciences et de Nanotechnologies, Université Paris-Saclay, 91405 Orsay, France
3. Laboratoire Lumière, Matière et Interfaces, Université Paris-Saclay, 91405 Orsay, France

**Abstract**

We have studied nucleation of magnetic domains and propagation of magnetic domain walls (DWs) induced by pulsed magnetic field in a ferromagnetic film with in-plane uniaxial anisotropy. Different from what have been seen up to now in out-of-plane anisotropy films, the nucleated domains have a rectangular shape in which a pair of the opposite sides are perfectly linear DWs, while the other pair present zigzags. This can be explained by magnetostatic optimization, knowing that the pulse field is applied parallel to the easy magnetization axis. The field induced propagation of these two DW types are very different. The linear ones follow a creep law identical to what is usually observed in out-of-plane films, when the velocity of zigzag DWs depends linearly on the applied field amplitude down to very low field. This most unusual feature can be explained by the shape of the DW, which makes it possible to go round the pinning defects. Thanks to that, it seems that propagation of zigzag walls agrees with the 1D model, and these results provide a first experimental evidence of the 1D model relevance in two dimensional ferromagnetic thin films. Let's note that it is the effective DW width parallel to DW propagation direction that matters in the 1D model formula, which is a relevant change when dealing with zigzag DWs.

Keywords: in-plane magnetic thin film, magnetic domains structure, domain wall motion, 1D model.

**Introduction**

Propagation of magnetic domain wall (DW) has been widely studied for out-of-plane magnetic thin films[1-6]. In these kind of films, without the symmetry break induced by phenomena such as Dzyaloshinskii-Moriya interactions[7-11], DW propagation is highly isotropic. As a result, when the applied field is sufficient to overcome the pinning effects, the shape of magnetic domains is expected to become circular, and quite amazing circular domains have been indeed observed[3,4].

Now, out-of-plane magnetic films are not the only ones, and quite surprisingly, up to now, there are very few works about DW propagation in in-plane full films. The main difficulty is the very fast DW velocity of such kind of samples and the quite small view field allowed by longitudinal microscope. So, except for GaMnAs, for which it has been possible to carry out an interesting study of the field dependence of DW velocity[12], one has to use tricks to get some dynamic data: some authors have tried to use magnetic relaxation[13], others have used nanostructured wires and a laser spot[14]: by analyzing the signal, velocity as a function of the magnetic field could be deduced. But, these samples can be viewed as 1D sample, DW extends over the whole width of the wire and can be considered as rigid. In 2D films, it is no more true, DW can bend and go around a pinning defect. So, the dynamic behavior in in-plane magnetic thin films is really poorly known and experiments using Kerr microscopy have been mainly focused to the study of static configuration.

Using high amplitude short field pulses, how nucleation occurs in in-plane films? What can we expect for the propagation? What domain shape will we get?

Here, we present a study of the dynamic behavior of DW movement in in-plane magnetic films using very short field pulses. Our set-up could create field pulses of length as short as 1 µs and of amplitude up to 2 mT, making it possible to reach the fast velocity regime. After showing the experimental results, they are analyzed in the framework of the 1D model.

**Samples and experiments**

*1. The fabrication of the samples and the basic magnetic properties*

We have mainly studied the film of Si/Ta(2 nm)/CoFeB(30 nm)/Ta(1 nm). All the results presented here have been obtained with this stack. The film in study was grown at 300 K on the Si (100) substrate by a high vacuum dc sputtering system. During the growing of the film, there was an in-plane magnetic field of around 1 mT applied to induce an uniaxial anisotropy in the film. The target material is $Co_{60}Fe_{20}B_{20}$. In order to protect the magnetic layer from oxidation, a Ta layer was sputtered above the magnetic film.

Several preliminary experiments have been performed to determine the magnetic properties of the film. Magnetization was measured by a vibrating sample magnetometer (VSM) and the saturation magnetization $M_s$ was found equal to $9.6\times10^5$ A.m$^{-1}$. In-plane Kerr hysteresis loop using longitudinal magneto-optic Kerr effect (LMOKE) has enabled us to check the coercive field as well as the anisotropy. The easy axis (EA) and hard axis (HA) have been identified (see Fig. 1(a)) by checking the hysteresis loops as a function of the angle between the long edge of sample and the applied magnetic field. Coercive field has been determined equal to 0.40 mT when magnetic is along the EA. Moreover, Kerr loop measured along HA was used to get the in-plane magnetic anisotropy (see Fig. 1(a)). The shape of the loop agrees with an anisotropy energy $E_a = -K_a M_s^2 Cos^2\theta$, where θ is the angle between the magnetization and the EA[15-19]. The anisotropy field $\mu_0 H_k = 2K_a/M_s$ has been found equal to 3.5 mT, which is in perfect agreement with the result of complementary ferromagnetic resonance experiment (see the first part in SI) by which we could also get the Gilbert damping, α with the value equal to 0.0085 and error within 5%.

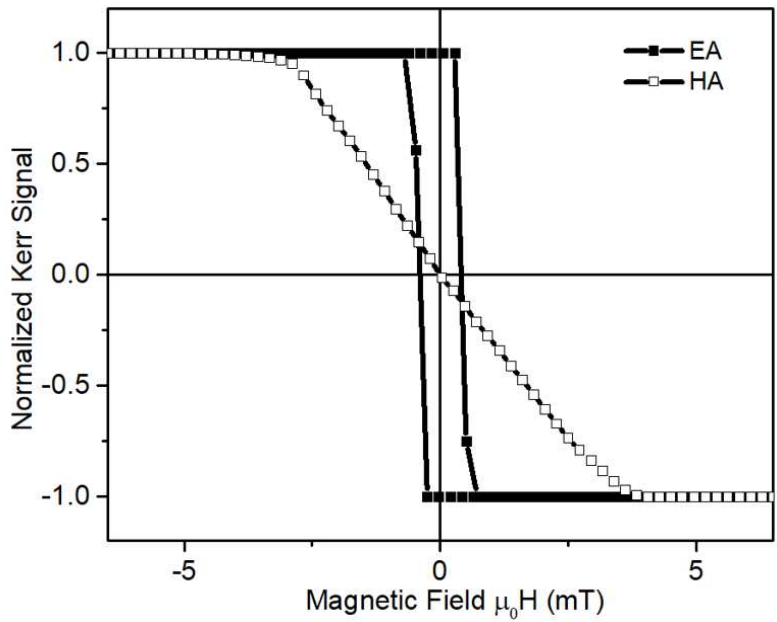

FIG.1 The Kerr hysteresis loops for magnetic field parallel to easy magnetization axis (EA) and hard magnetization axis (HA).

*2. Method for the measurement of magnetic domain wall velocity*

DW motion was investigated by a longitudinal Kerr microscope at room temperature[20-23]. In this setup, a parallel polarized light beam was arriving on the sample with an incidence angle of 45°, giving rise to a quite big longitudinal Kerr rotation. The reflected beam was focused on a CMOS camera, the CMOS sensor plane and the objective were slightly tilted with respect to the beam propagation axis[20,23], so that image plane of the film was in the plane of the sensor. Optical resolution of this setup was around 30 µm.

The coplanar magnetic pulse field was produced by a small coil of radius 17.5 mm centered on the sample. The field-of-view of the microscope was less than 10 mm, so that the field created by the coil was uniform within a precision of 2% over the area studied. The inductance of the coil was between 4 µH and 40 µH depending on the coil used, making it possible to create very short field pulses. A high voltage pulse generator was used, so that, with the coil in serial with a resistive charge of 50 Ω, we could get current pulses as big as 15 A, corresponding to a magnetic field of 2 mT. The best risetime with the lowest inductance coil was 83 ns, making it possible to have pulses as short as 1 µs. The sample holder was made of plastic, i.e. insulating material, so that there was no eddy current modifying the characteristics of the magnetic field generated by the coil. For the work presented here, the magnetic field was always applied along the EA of the sample. DW Velocity was measured using the usual stroboscopic way and Kerr microscopy[1,2,12] (see also SI).

**Results and discussion**

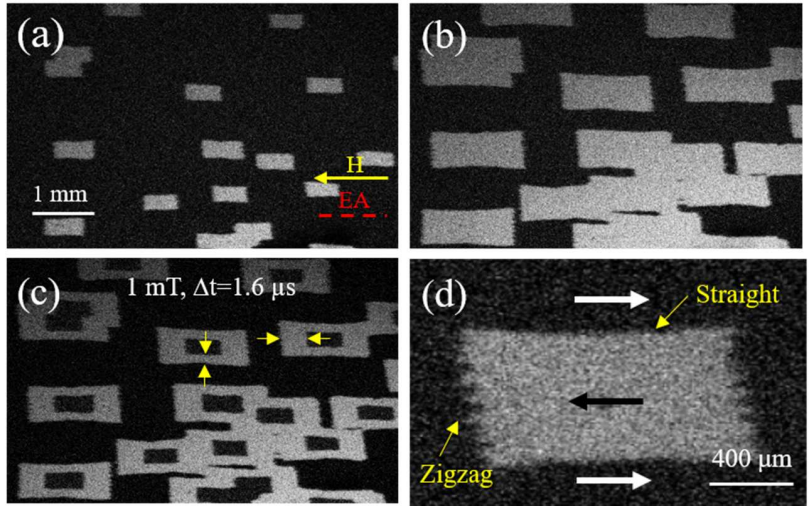

FIG. 2 shows a typical sequence to measure velocity when the length of the pulsed field is much longer than the risetime. Starting from a saturated state, figures (a) and (b) show the full-view Kerr images after the application of the (a) first and (b) second pulse field. The pulsed magnetic field (yellow arrow) was parallel to EA (red dash) with amplitude of 1 mT and length of 1.6 μs. (c) shows the DW displacement during the second pulse Δt=1.6 μs. (d) presents the magnification of a rectangle domain with two types of DW (horizontal straight and vertical zigzag) being indicated. White (black) arrows denote the magnetization directions outside (inside) the domain.

A typical example of DW motion is shown in Fig. 2 in which (a) shows the nucleation and (b) shows the domains after propagation due to the second pulse. Fig. 2(c) shows the difference between these two pictures, making it easier to measure the motion length.

The first important result is the shape of the nucleated domains (Fig. 2(a)), as well as the shape after pure propagation (Fig. 2(b)): in both cases, we got the same highly anisotropic rectangular form. It is true that magnetostatic optimization leads to rectangular domains[23,24], but, it was not obvious that, using very short pulses, we would not have a different metastable shape. In addition, we can see that the limiting DW on the horizontal part is almost straight, when the ones on the vertical sides of the rectangle present zigzag structures (Fig. 2(d)). This can be explained by static energy optimization[23,25-28]: on these sides, because of the in-plane anisotropy, a straight vertical DW would mean head to head DW (also called charged DW). To avoid the high energetic cost due to this kind of DW, zigzags appear. It increases the length of the DW, which increases energy, but, the energetic decrease due to a much less charged DW compensates it. Here, we have found that the zigzag angle θ of the zigzag DW (see Fig. 3(a)) did not depend on the amplitude of external pulse field and its value has been found around 22° (±1.5°).

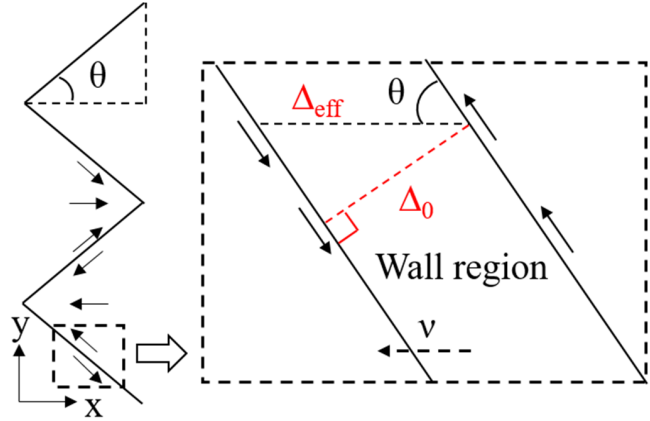

FIG. 3 Left panel: schematic of a zigzag wall as well as the definition of geometrical zigzag angle θ; Right panel: magnification of a single segment of zigzag wall with the intrinsic DW width $\Delta_0$ and effective DW width $\Delta_{eff}$ indicated by red and black dashed lines respectively. Note that the effective DW width was defined parallel to the propagating direction of the DW marked by dashed arrow.

The second important result is the velocity: it is not the same for horizontal straight DW and vertical zigzag DW, as can be seen in Fig. 3(c). Velocities for the two DW types have been plotted in Fig. 4. First, vertical zigzag DWs go faster than horizontal straight DW. Second, the corresponding dependence of velocity as the function of the applied field is quite different: for horizontal straight DWs, we have a non-linear law, which agrees with the creep law usually observed in out-of-plane thin films[1,2]:

$$v(H) = v_0 \cdot \mathrm{Exp}\left[\left(\frac{H_p}{H}\right)^{\frac{1}{4}}\right] \qquad (1)$$

The velocity range extends over almost 2 orders of magnitude, which becomes meaningful to validate the creep behavior. As a matter of fact, pinning effect is quite common in any magnetic films due to the inevitable imperfection, and it seems logic to get it here in the in-plane films. However, it becomes extremely different when we look at the zigzag DW: here, we got a very unusual and impressing linear law, which really goes through zero when the applied field goes to zero. Such a behavior has been predicted by the 1D model[29] which however assumes a perfect film without any pinning effects.

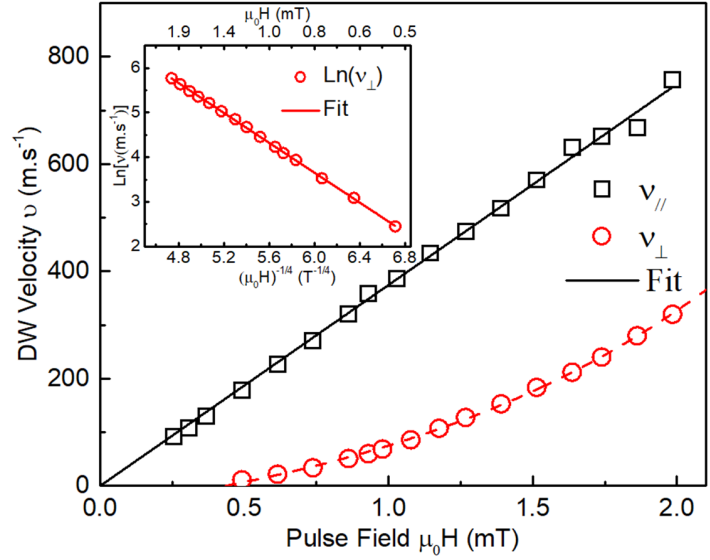

*FIG. 4 DW velocity as the function of pulse field for both zigzag and straight wall denoted by v// (black open square) and v⊥ (red open circle) respectively. The black solid line is the linear fit with the formula v=μ·H. Red dash is the guide for eyes. The insert shows the plot of Ln(v) vs. H$^{-1/4}$ for the straight wall with the linear fit (red solid line) using Eq. (1).*

**Analysis**

Because we have a perfect linear law for zigzag walls, we have assumed that 1D model applies. According to it, for small magnetic field below the Walker breakdown, we should have the following velocity law:

$$v(H) = \frac{\gamma \mu_0 H \Delta}{\alpha} \tag{2}$$

where α is the Gilbert damping parameter, γ the gyromagnetic ratio, $\mu_0$ the vacuum permittivity, H the applied field and Δ the width of the DW, assuming a DW profile[29,30]:

$$\varphi = 2 Arctan\left[\frac{Exp(x - x_0)}{\Delta}\right] \tag{3}$$

From this, we can check Δ. From the fit of Fig. 3(a), the slope, we have found Δ = 17 nm, which means a real width πΔ = 53 nm, this is a quite small value for an in-plane film, but it is the right order of magnitude[23,24] and this is one more point for the validity of 1D model. Let's note that we have not been able to view the Walker breakdown: using the 3D threshold $\alpha M_s/2$[29], we have found a typical value of 6 mT. Because our sample was not 3D but rather 2D, the Walker breakdown might be different[31]. However, from this estimated value, we can think we did not reached the critical Walker value.

Now, the question is: why 1D model applies for zigzag DW when it obviously doesn't for horizontal straight DW? A possible explanation is that zigzag is a way of inhibiting the pinning defect effect. Indeed, if the DW meets a pinning defect, it remains possible for it to go on by putting a zigzag angle at this very point and, as it is going on, the amplitude of the zigzag is just increased. According to that, for low field, the main pinning point might be hard to overcome, it could explain fewer and irregular zigzag (see SI, Figure S5). Above a threshold field, pinning points becomes negligible, and the zigzag density becomes constant and equal the magnetostatic equilibrium value.

Now, a last question is how should we measure $\Delta$? Indeed, for the horizontal straight DW, the width seems obvious, but the vertical zigzag ones? Should we use $\Delta_{eff}$ or $\Delta$ (see Fig. 3)? When calculating the velocity, one assumes $\varphi(x - x_0(t))$, where $\varphi$ is the tilt angle of the local magnetization, going from 0° to 180° along the wall and $x_0$ the position of the wall[24,29]. Through this method, the propagation velocity $v = dx_0/dt$ is linked to $\partial\varphi/\partial t$ through $\partial\varphi/\partial t = -v\, d\varphi/dx$. As $d\varphi/dx$ is proportional to $1/\Delta_{eff} = \mathrm{Sin}\theta/\Delta_0$, we expect DW velocity to be proportional to $\Delta_0/\mathrm{Sin}\theta$, where $\Delta_0$ is the "intrinsic width" measured perpendicularly to the DW direction.

Note that the x axis can be chosen in any direction, it doesn't matter: for an infinite straight DW, the final result is the same. Indeed, if you translate such a DW over $\Delta_0/\mathrm{Sin}\theta$ in the x direction, whatever is the x direction, starting from the same initial position, one gets the same final position (see Fig. 3(a)).

To check that, using optical lithography, we have patterned wires from one of our 30 nm thick CoFeB samples (see Fig. 5(a)). The wires were narrow enough to avoid possible zigzag within their width. Several sets of wires with different directions were patterned to check the effect of the in-plane anisotropy. Velocity was checked at quite high field, so that pinning had become negligible and a 1D behavior according to formula (3) could be expected. Fig. 5(b) and (c) show the nucleation and propagation on one set of wires. Quite surprisingly, the angle θ was not the same for all wires and didn't seem to depend on the anisotropy axis of the film. This maybe resulted from that the annealing during the patterning process had destroyed the EA. Thanks to this, we could plot the mobility $v/\mu_0 H$ as a function of $1/\mathrm{Sin}\theta$: as expected, a very nice linear law has been obtained and plotted in Fig. 5(d). In addition, in this graph, we have added in red the point for the zigzag wall on the full film using the zigzag angle 22° obtained above: this point aligns very well with the other ones.

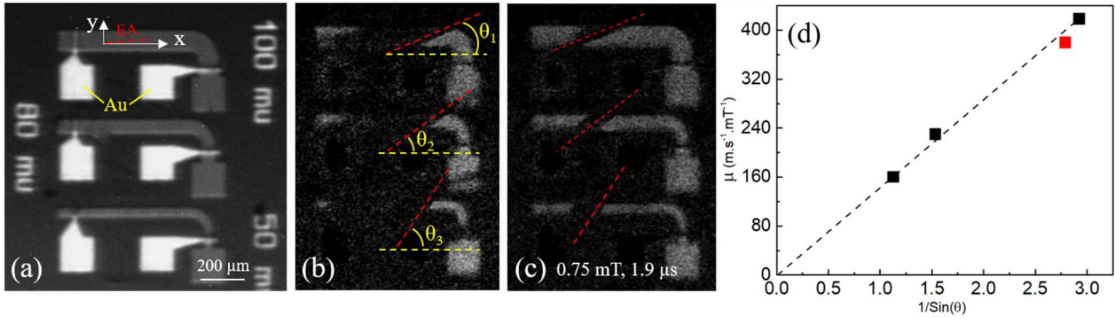

FIG. 5 (a) An optical image of the L-shaped microwires of Ta(2 nm)/CoFeB(30 nm)/Ta(1 nm) stack with wire width of 100, 80, and 50 µm (from top to bottom). The EA has been marked by a red dashed line. The white parts are Au electrodes deposited on the top of the wires (not used in this work). (b) The initial DWs state in which DWs with the 'slant' angle $\theta_1$, $\theta_2$ and $\theta_3$ nucleated with a certain pulse in 100, 80 and 50 µm wires respectively. (c) A typical Kerr image of DWs in the wires after the application of a field pulse with amplitude of 0.75 mT and length of 1.9 µs. (d) The measured DW mobility µ as the function of $1/\mathrm{Sin}(\theta)$ for the three 'slant' DWs in the wires. The DW mobility for zigzag wall measured in the full film has also been displayed by the red square.

**Conclusion**

In conclusion, we have found a highly anisotropic dynamical behavior in an in-plane magnetic thin film of Ta/CoFeB/Ta. Using magnetic field pulses parallel to the easy plane of our film, the shape of the domains nucleated by a pulse was a rectangle. The limiting DWs of these rectangular domains were different according to the sides. The two sides parallel to the easy magnetization axis were straight lines, while the other two sides presented zigzag, as predicted by magnetostatics to avoid charged DWs. Depending on the type of sides, the propagation velocity was very different: on the side along the easy axis, where DW was almost a straight line, the velocity seems to follow a creep law. But, on the other sides, the zigzag DW presented a impressing linear law, in good agreement with the 1D model. We suggest that the possibility of creating zigzag at the blocking defects destroys the effect of the pinning. At last, we have pointed out that the velocity is also changed because of the tilting induced by the zigzag. We have shown that velocity is in fact proportional to the effective DW width, i.e. the width obtained when measuring it along the propagation direction. Let's add that some preliminary results with a permalloy film shows that the behavior seems to be the same : our results require only in-plane anisotropy with an easy axis in the plane to occur.

This work rises many new open questions: the effective width found for our DW seems quite small for an in-plane film, why? Can we find the Walker breakdown in full in-plane film? In the framework of this work, we have stuck to the parallel case in which magnetic field was applied along the easy magnetization axis, what would happen if the magnetic field was applied in another direction?

**Acknowledgements**

This work is supported by the National Key Research and Development Program of China (Grant No. 2017YFA0204800), NSFC (Nos. 51571062) and China Scholarship Council for a grant. The authors wish to thank André Thiaville for his useful advices.

Supplementary informations

Highly anisotropic magnetic domain wall behavior in-plane magnetic films

Xiaochao ZHOU[1,2], Nicolas VERNIER[2,3], Guillaume AGNUS[2], Sylvain EIMER[2], Ya ZHAI[1]

1. School of Physics, Southeast University, 211189 Nanjing, China
2. Centre de Nanosciences et de Nanotechnologies, Université Paris-Sud, 91405 Orsay, France
3. Laboratoire Lumière, Matière et Interfaces, Université Paris-Saclay, 91405 Orsay, France

**) Ferromagnetic resonance:**

Figure S1(b) shows the angular variation of the resonance field in the film studied which exhibits a pure uniaxial feature without fourfold anisotropy, i.e. $K_4$ is negligible in this system. Given the experimental data of resonance fields $H_{R1}$ and $H_{R2}$ at Easy Axis (EA) and Hard Axis (HA) respectively, by solving Kittel equation the in-plane anisotropy field $H_a$ was found approximately $H_a \sim (H_{R2} - H_{R1})/2$ as indicated in Figure S1(b). Details has been shown below.

Starting from the free energy of the studied system in the spherical coordinate (see Figure S1(a) for the definition):

$$F = -\mu_0 H M_s Sin(\theta) Cos(\varphi - \varphi_H) + K_p Cos^2(\theta) + K_a Sin^2(\theta) Sin^2(\varphi).$$

In which θ and φ are the polar and azimuthal angles of the magnetization vector respectively and $\varphi_H$ is the azimuthal angle of the external magnetic field with respect to EA; $M_s$ is the saturation magnetization; H is the applied external field; $K_p = \left(K_\perp - \frac{1}{2}\mu_0 M_s^2\right)$ is the sum of the perpendicular anisotropy term and demagnetizing term representing the total anisotropy field of out-of-plane symmetry. $K_a$ is the in-plane uniaxial anisotropy constant. According to Kittle equation, ferromagnetic resonance occurs at a free energy given by:

$$\left(\frac{\omega}{\gamma}\right)^2 = \frac{1}{M_s^2 Sin^2(\theta)} \left[\frac{\partial^2 F}{\partial \theta^2}\frac{\partial^2 F}{\partial \varphi^2} - \left(\frac{\partial^2 F}{\partial \theta \partial \varphi}\right)^2\right]$$

$$= \mu_0^2 [H_R Cos(\varphi - \varphi_H) + H_p + H_a Cos^2(\varphi)] \cdot [H_R Cos(\varphi - \varphi_H) + H_a Cos(2\varphi)].$$

Where $H_R$ is the resonance field, $H_p = 2K_p/M_s$ and $H_a = 2K_a/M_s$ are the out-of-plane anisotropy field and the in-plane uniaxial anisotropy field respectively. Taking into consideration the experimental data at EA ($\varphi=\varphi_H=0$) and HA ($\varphi=\varphi_H=\pi/2$), and the fact that $K_p \gg K_a$. $H_a$ can be deduced approximately by $H_a \sim (H_{R2} - H_{R1})/2$ in which $H_{R1}$ and $H_{R2}$ are resonance fields measured at EA and HA respectively.

The value of $H_a$ corresponds to the fitting line (solid line in Figure S1(b)) was obtained 3.6 mT, in a good agreement with the value obtained from hysteresis loop. Gilbert damping α was also obtained by FMR which read 0.0085 with error within 5%.

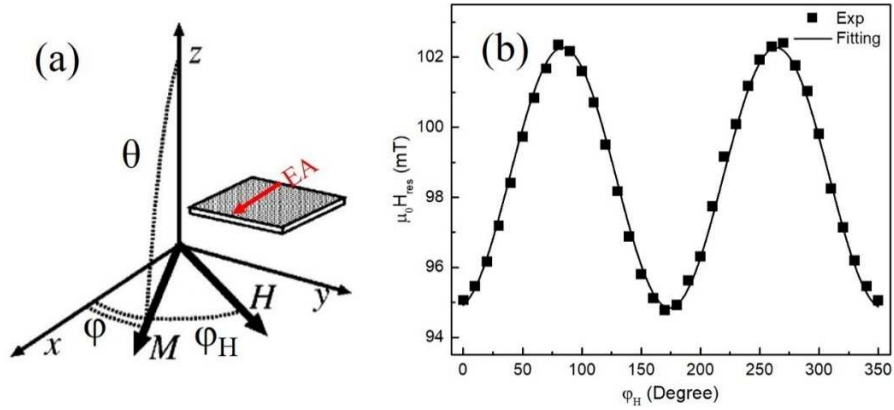

*Figure S 1 (a) Schematic of the coordinate system in which x axis was defined along the EA (red arrow). (b) Ferromagnetic resonance field for Ta(2 nm)/CoFeB(30 nm)/Ta(1 nm) film (black square) as a function of in-plane angle $\varphi_H$ with respect to EA. The solid line is the fit based on Kittle formula.*

) **Shape of the magnetic pulses**

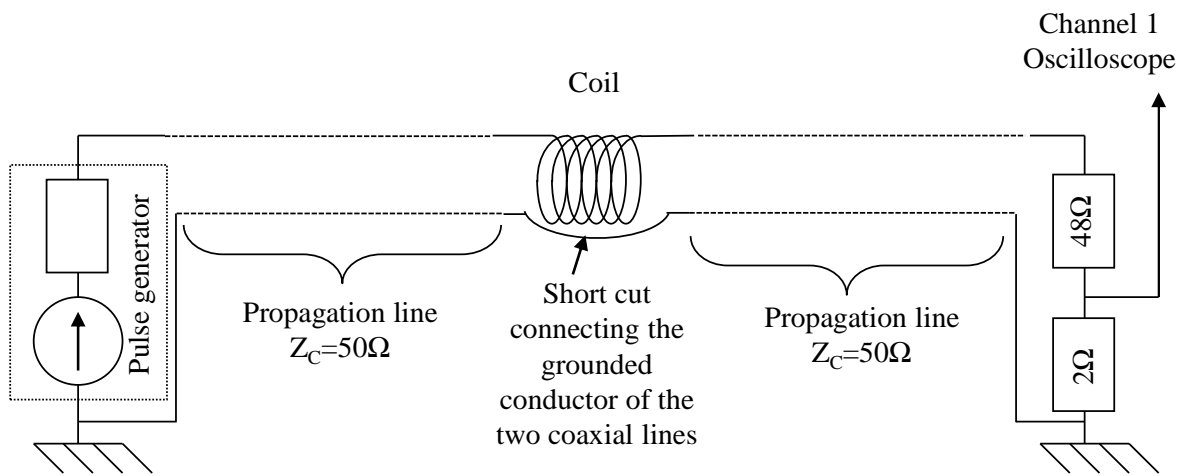

*Figure S 2 Schematic of the pulse field circuit.*

The coil was connected the following way: the inner conductor of the coaxial line leaving the generator was connected to the first wire of the coil. The second wire was connected to the inner conductor of a second coaxial line. The ground of both cables was short-cut with a very small cable. The other end of the second coaxial line arrived on 50 Ω ending, made with a 48 Ω resistor in serial with a 2 Ω resistor, the latter one being the one connected to the ground. An oscilloscope was connected in parallel with the 2 Ω resistor, so that, we have a voltage divider and the voltage detected by the oscilloscope was reduced to no more than 30 V, aiming to protect the electric circuit. The ending resistors were put in a very small connecting box directly put on the oscilloscope, so that no propagation phenomenon could modify the signal. As the current going inside the coil has to go through the second coaxial line and through all the 50 Ω ending, the voltage detected by the oscilloscope was directly proportional to the current flowing in the coil and to the magnetic field created. So that, we were able to check real time the magnetic field applied and the shape of the magnetic pulse. Let's note that at the input of the coil, there is an impedance mismatch. However, as the size the coil is small, the mismatch impedance is a small one (as bandwidth required is a few 10 MHz. Using 100 MHz, we get a minimum wavelength of 3 m, which is indeed much bigger than the size of the coil). So, the reflection is also very small and the current detected is anyway the one going inside the coil.

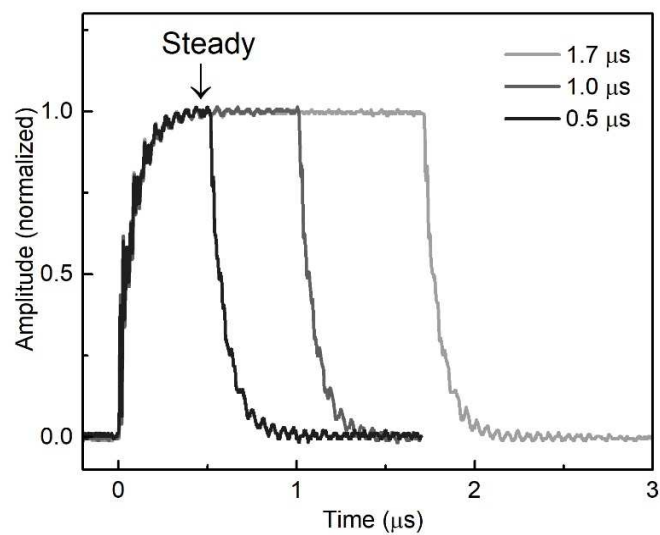

*Figure S 3 The shape of the current pulses with the length of 0.5, 1.0 and 1.7 μs generated in the coil.*

) **Method for velocity measurement**

In this work, the voltage pulse generator combined with our home-made coil can provide pulsed field with amplitude up to 2 mT. As the pulsed field increases from 0 to 2 mT, three different kinds of magnetic domain nucleation and propagation have been identified in different field regions namely the low field region H<0.6 mT, the intermediate region 0.6<H<1.1 mT and the high field region H>1.1 mT. Two methods of measuring DW displacement have been utilized accordingly and introduced respectively as below.

1) Measurement of DW displacement in low and high field region.

In high field region, rectangular magnetic domains nucleate at the center area of the sample surface. Thanks to the regular shape, DW displacement along x and y directions can be easily measured as described below. To begin with, the sample was saturated in one direction along EA using a long pulse of magnetic field of high amplitude, a reference picture was then acquired. Next, a first magnetic field pulse with proper amplitude applied in the opposite direction nucleated several domains on the sample surface. A second picture was then acquired. To finish, a second pulse of desired amplitude was applied and a third picture was acquired. The difference between the second and the third pulses was the DW displacement during the last pulse and the velocity can be deduced. This procedure was used when the risetime was negligible as compared to the duration of the magnetic field pulse. A typical example of this procedure was shown in Fig. 2 in the manuscript. For very short pulses (< 1 µs), in order to remove the transient movement occurring during the rise time and keep only the movement happening in the required steady field (see steady field in Figure S3), a set of two pulses was used after nucleation. First, a short pulse of duration $t_1$ was applied. A picture was then acquired and a first displacement $d_1$ during this very short pulse has been obtained. Then, a shorter pulse of duration $t_2$, just long enough to reach steady field (≥ 0.4 µs in this study, see Figure S3) was applied, a new picture was acquired and the displacement $d_2$ during this second pulse was measured. As the transient displacements occurring during the rise and the decay times for these two pulses were the same, the true steady state velocity can be obtained through the formula $(d_1-d_2)/(t_1-t_2)$. A typical result of this procedure was shown in Figure S4 with $t_1$, $t_2$, $d_1$ and $d_2$ being indicated. The result image was obtained by operating add function between two Kerr images captured after the above mentioned two pulses, by which the DW displacement occurred within each pulse can be well distinguished due to the contrast. In this case, the dark (bright) area was the displacement occurred in the first (second) pulse with amplitude 1.75 mT and length $t_1=0.7$ µs ($t_2=0.4$ µs) and the corresponding horizontal displacement $d_1$ ($d_2$) was marked by yellow (orange) arrows as shown in the figure. By this procedure, the DW velocities for both the vertical zigzag wall and horizontal straight wall can be obtained.

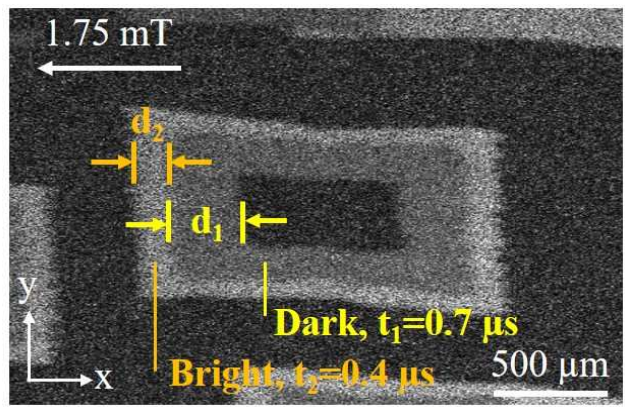

*Figure S 4 Successive DW displacements, the dark and bright areas were driven by two short pulses with pulse length of $t_1=0.7$ and $t_2=0.4$ µs respectively. The pulse amplitude was 1.75 mT and direction were indicated by white arrow. The horizontal displacement $d_1$ and $d_2$ occurred in each pulse were denoted by yellow and orange arrows respectively.*

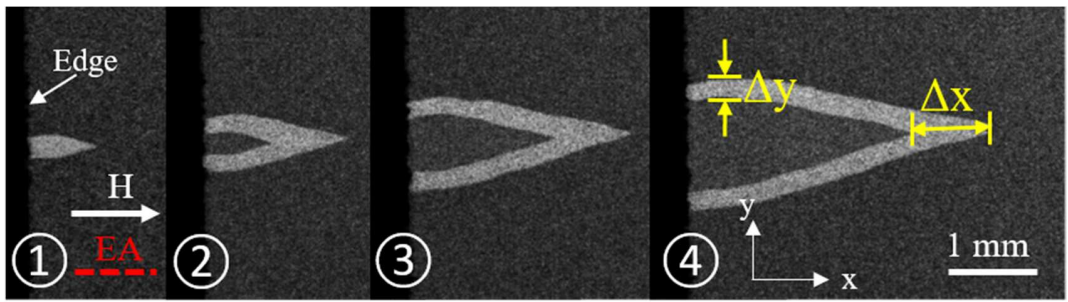

*Figure S 5 ① is the Kerr image of a magnetic domain nucleated at the edge of sample after the first field pulse and ②-④ are the successive DW displacements in the following three pulses respectively. The applied field pulse was set parallel to the EA with amplitude of 0.40 mT and length of 8 µs. The way of measuring the DW displacements along x (parallel to the EA) and y (perpendicular to the EA) direction has been depicted in ④.*

Note that in low field region, magnetic domain(s) only nucleated in several points at the edge of the sample and then extended towards the center. A typical example of this process has been shown in Figure S5 as well as the illustration of the way to determine DW displacement along x and y direction. By this way, the average velocity along and perpendicular to the EA can be obtained.

2) Measurement of DW displacement in intermediate region.

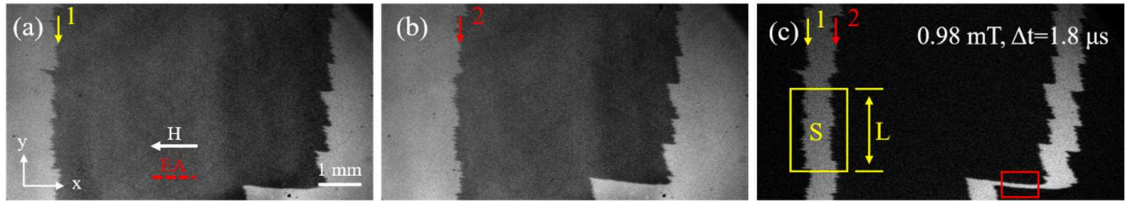

*Figure S 6 Kerr images of the magnetic domains (bright areas) before (a) and after (b) the application of a field pulse with amplitude of 0.98 mT and length of 1.8 μs. The front line of the left zigzag DW before and after the applied field pulse is denoted by 1 and 2 respectively. (c) is the DW displacements (bright areas) occurred in the pulse time. Yellow frame is the selected long segment with length of L for calculating the average DW displacement along x direction. DW displacement along y direction is measured by the usual method in the red frame.*

In the intermediate field region, magnetic domain nucleated everywhere at the sample edge forming a zigzag DW and propagated towards the center. Figure S6 shows a typical example. In this case, the front line of the left zigzag DW marked by 1 propagated along x direction to the final positon marked by 2 in the Kerr images in Figure S6(a) and (b) and the DW displacement happened in a pulse time has been shown by the bright area in (c). As the zigzag DW usually presented an irregular front line in this field region, The average DW displacement along x direction was then obtained by S/L where S is the area of the bright area in the selected long segment with the length of L marked by the yellow frame in Figure S6(c). Thus, the average velocity of zigzag DW along x direction $v_{\parallel}$ can be deduced. Note that the DW displacement along y direction and the corresponding vertical velocity $v_{\perp}$ in this case was measured by the similar way used in the low/high field case introduced above as the displacement of the straight DW is quite uniform indicated in the red frame in Figure S6(c).

) **Influence of magnetic pinning on a field-driven zigzag DW**

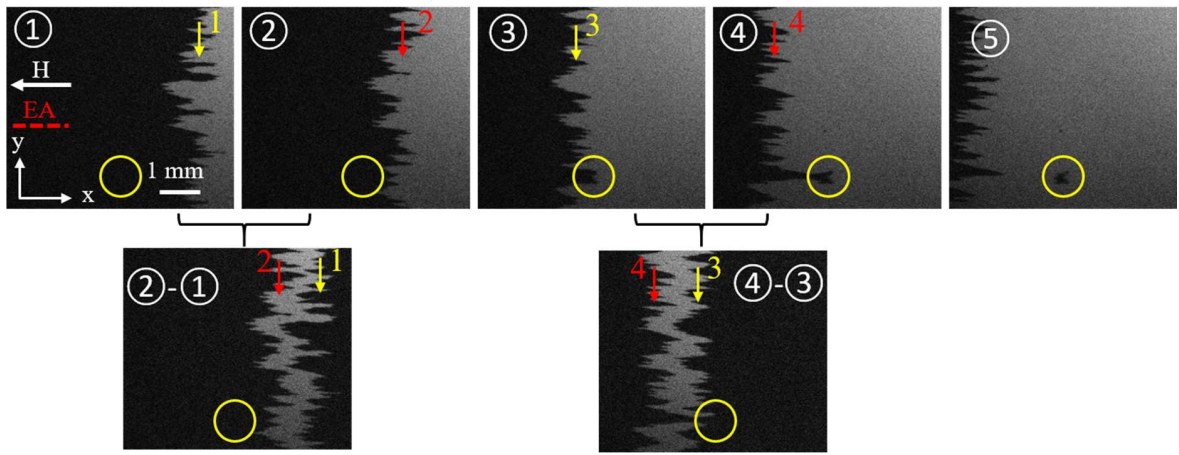

*Figure S 7 Top line: Propagation of a zigzag DW nucleated at the right edge of the sample under the successive application of 4 field pulses with amplitude of 0.62 mT and length of 4 µs. Yellow circle in each Kerr image indicates the strong pinning site. Number 1-4 are the indicator of the front line of zigzag DW in the Kerr images. Bottom line: DW displacement happened during the first (third) pulse before (after) the DW meeting the pinning site.*

Lattice defects and grain boundaries form numerous discrete magnetic pinning sites in soft magnetic films and nanostructures which has long been considered as the main reason that DW moves in the way of Barkhausen jump leading to the observed creep feature in v-H curve. However, this effect seems to be effectively inhibited for zigzag DW of which the velocity has been investigated in this work to exhibit a linear dependence with the field amplitude. This is in a good agreement with the prediction of 1D model in which no magnetic pinning is considered (see Fig. 4 in the full manuscript). An evidence can be found in Figure S7 which shows a typical process of how a zigzag DW passes a strong pinning site indicated by the yellow circle in each Kerr image. As can be seen, zigzag DW leaves a zig pinned at the strong pining site while moving past it. Keep injecting the same field pulse, the front line of zigzag DW keeps moving forward with the pinned zig getting deeper and without losing DW velocity evidenced by the approximately equal DW displacement along x direction in the same pulse duration measured before and after meeting the pinning site (shown in the bottom line of Figure S7). In detail, the average displacements measured by the method provided in the section □ before and after meeting the pining site have a difference within 5%. Thus, we can conclude that the velocity of zigzag DW maintains no change while passing pinning sites.

On the other hand, as the pulse amplitude increases, it's getting easier to move past the pinning sites for a zigzag DW as evidenced by the smaller depth of the pinned zigs shown in Figure S8. In fact, when the pulse amplitude is beyond 1.0 mT, pining sites almost have no impacts to the zigzag DW anymore and the zigzag properties like zigzag amplitude and density are only determined by the magnetostatic interaction in the film.

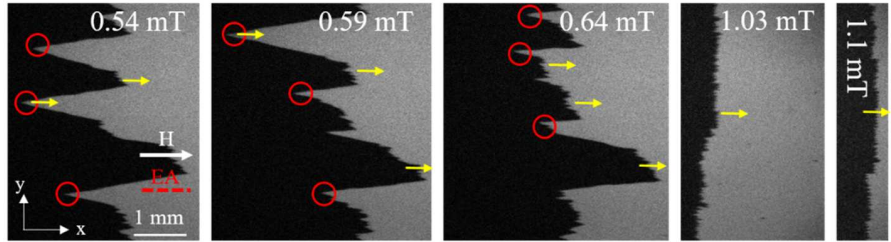

*Figure S 8 Zigzag DW of the magnetic domain (dark area) nucleated at the left edge of the sample after the application of several field pulses with amplitude of 0.54, 0.59, 0.64, 1.03, and 1.10 mT respectively. Red circles denote several pinning sites in the film. Yellow arrows denote the propagation direction of zigzag DW.*